\documentstyle[aps, prb, epsfig]{revtex}
\newcommand{\figdir}{./}
\newcommand{\pins}{p_{I}}
\newcommand{\psc}{p_{S}}
\newcommand{\sins}{\sigma_{I}}
\newcommand{\ssc}{\sigma_{S}}

\newcommand{\spar}{\sigma_{{\small M},\|}}
\newcommand{\sper}{\sigma_{M,\perp}}

\newcommand{\beq}{\begin{equation}}
\newcommand{\unbeq}{\end{equation}}
\newcommand{\beqa}{\begin{eqnarray}}
\newcommand{\unbeqa}{\end{eqnarray}}

\begin{document}
\twocolumn[\hsize\textwidth\columnwidth\hsize\csname
@twocolumnfalse\endcsname
\title{
Magnetoresistance of Three-Constituent Composites:
Percolation Near a Critical Line
}
\author{Sergey V.\ Barabash$^1$, David J.\ Bergman$^{1,2}$, D.\ Stroud$^1$}
\address{$^1$Department of Physics,
The Ohio State University, Columbus, Ohio 43210;\\
$^2$School of Physics and Astronomy, Raymond and Beverly Sackler
Faculty of Exact Sciences,\\Tel Aviv University, 
Tel Aviv 69978, Israel}

\date{\today}

\maketitle

\begin{abstract}
Scaling theory, duality symmetry, and numerical simulations of a random
network model are used to study the magnetoresistance of a 
metal/insulator/perfect conductor composite with a
disordered columnar microstructure.
The phase diagram is found to have a critical line which separates regions 
of saturating and non-saturating magnetoresistance.  The percolation
problem which describes this line is a generalization of anisotropic
percolation.  We locate the percolation threshold and determine the
values of the critical exponents
$t_{\|} = t_{\perp} = s_{\|} = s_{\perp} = 1.30 \pm 0.02$, 
$\nu=4/3 \pm 0.02$, which are the same as in two-constituent 2D
isotropic percolation.
We also determine the exponents which characterize the critical
dependence on magnetic field, and confirm numerically that $\nu$ is
independent of anisotropy.
We propose and test a complete scaling description of the
magnetoresistance in the vicinity of the critical line.
\end{abstract}
\vskip1.5pc] 

\section{Introduction}

Magnetotransport in composite conductors has attracted
increased attention due to the discovery of new effects, like
the appearance of non-saturating magnetoresistance in a
metal/insulator columnar composite (denoted by $M/I$), induced
by the Hall effect in the $M$ (i.e., the metallic) constituent.
\cite{BS1} This means that, even when the $M$ constituent
has no {\em intrinsic} magnetoresistance but only a Hall resistivity,
an induced magnetoresistance appears in the mixture and continues
to increase as ${\bf B}^2$ ({\bf B} is the applied magnetic field)
{\em indefinitely} whenever the Hall-to-Ohmic resistivity ratio
of that constituent is much greater than 1. By contrast, in
a columnar composite of normal conductor and perfect
conductor (denoted by $M/S$), while there also appears an
induced magnetoresistance, it saturates when {\bf B} is
comparably large. \cite{BS1} Related new effects were also found in
periodic microstructures of either the $M/I$ type or the
$M/S$ type, where the induced
magnetoresistance often exhibits a strong anisotropy.
\cite{BergStrelPRB94,TornowWeissKlitzingEberlBergStrelPRL96}

In this paper, we study the magnetoresistance of 
three-constituent composites with a {\em random columnar microstructure}. 
Specifically, we analyze
a composite consisting of cylindrical $I$
and $S$ inclusions, though not necessarily circular-cylinders,
in an $M$ host film, with cylinder
axes perpendicular to the film, and with both a
magnetic field and a uniform macroscopic or volume averaged electric current
applied {\em in the film plane}.
We assume that the $M$ constituent has a
Hall effect, with a Hall-to-transverse-Ohmic resistivity ratio
$H \equiv \rho_{Hall}/\rho_{Ohmic} = \mu|{\bf B}|$,
where $\mu$ is the Hall mobility.
We are concerned with the effective resistivity tensor $\hat \rho_e$ of this
system at {\it large} $H$
($\hat \rho_e$ is
defined by $\langle {\bf E}\rangle = \hat \rho_e\langle {\bf J}\rangle$, where 
$\langle {\bf E}\rangle$ and $\langle {\bf J}\rangle$ are the volume averaged
electric field and current density.\cite{SSP})

This system might appear to be inherently three-dimensional (3D), 
because the Hall effect will generate an 
electric field with a component perpendicular to the film even with an
in-plane applied current. However, the electric field perpendicular to
the film plane vanishes because of the columnar $S$ inclusions.
Moreover, this 3D problem can be  
reduced\cite{BS1} to that of calculating the effective conductivity of 
a {\em two-dimensional} (2D)
composite of perfect insulator $I$ with $\sins = 0$, 
perfect conductor (or superconductor) $S$ with $\ssc=\infty$,
and anisotropic 2D metal $M$ with conductivity tensor
\beq
\label{2Dsigmas}
\hat{\sigma}_M
\equiv\left(
\begin{array}{cc}
\sper &  0   \\
  0   &\spar \\
\end{array}
\right)
= \frac{1}{\rho_M} \left( 
\begin{array}{cc}
\frac{1}{1+H^2}&  0 \\
  0  	      &   \frac{1}{\lambda} \\
\end{array} 
\right).
\unbeq
The conductivities $\spar$ and $\sper$  correspond to the
in-plane directions 
parallel and perpendicular to the applied magnetic field $\bf B$.
This transformation underlies our further discussion.
The form assumed for $\hat{\sigma}_M$ means that the $M$ constituent
remains an isotropic conductor, even in a magnetic field.
This excludes ``open orbit'' conductors, but includes the
possibility of some intrinsic magnetoresistance:
In that case both the transverse and the longitudinal Ohmic
resistivities, $\rho_M$ and $\lambda\rho_M$, would depend upon
$\bf B$. Our subsequent discussion will assume, for simplicity,
that $\lambda=1$ and $\rho_M$ are both independent of $\bf B$,
as would be the case if $M$ were a free electron or free hole
conductor. In that case, $H$ is simply proportional to $|{\bf B}|$.
The three-constituent
problem in 2D, but with a scalar $\hat{\sigma}_M$, was
treated previously in Ref.\ \onlinecite{kogut79}. Scaling and
simulational studies, of magnetotransport in 3D, two-constituent
$M/I$ composites with a random isotropic microstructure, were
also performed previously. \cite{SarBergStrelPRB93}

\section{Magnetoresistance and percolation}

\begin{figure}[b]
\centerline{ 
 \epsfig{file=\figdir 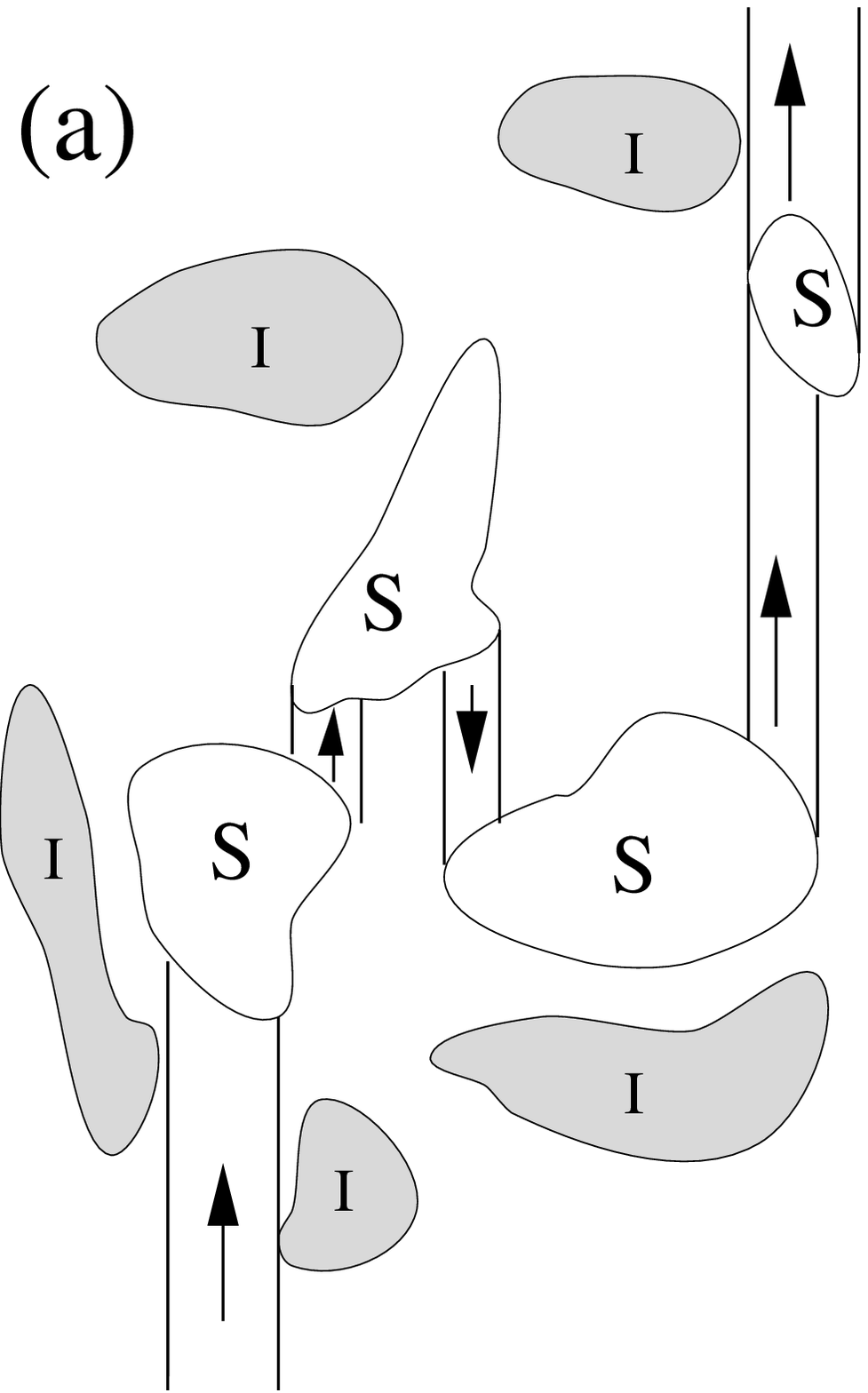, width=1.4in} 
 \epsfig{file=\figdir 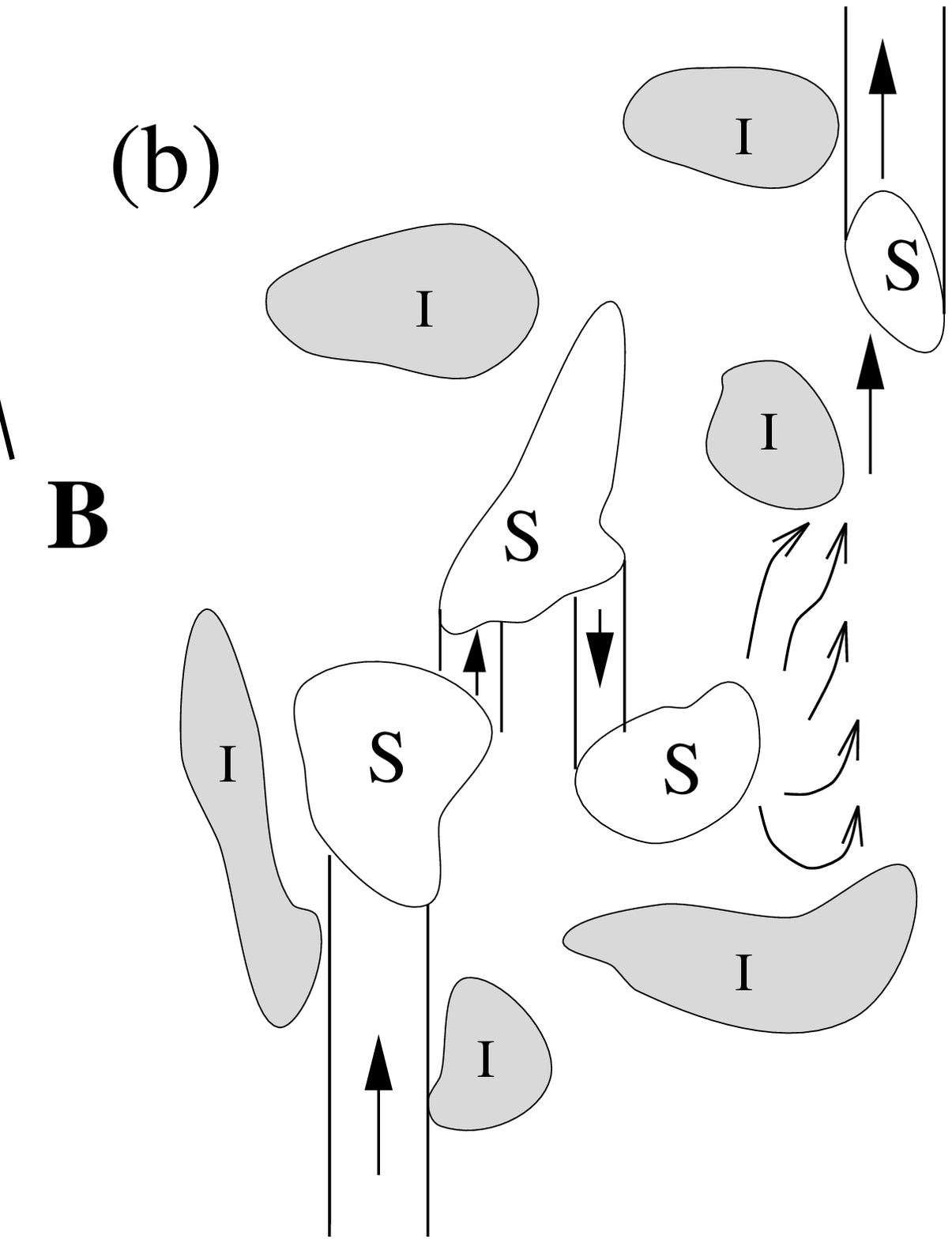, width=1.4in}
}
\vskip0.5pc
\caption[]
{
Schematic of transport in a 2D composite of a perfect conductor $S$,
insulator $I$, and a very anisotropic metal $M$
(the host medium in this figure).
The current prefers to flow between $S$ grains
only in the direction of high metallic conductivity $\spar$,
i.e., parallel to ${\bf B}$. 
At high $S$ volume fraction $\psc$, such a path always exists,
as shown schematically in (a).  At smaller $\psc$ or larger 
$\pins$, the current must sometimes flow through the
metal in the low-conductivity direction ($\sper\sim H^{-2}$), as shown in (b).
This behavior leads to a non-saturating effective resistivity.
}
\label{fig:percolation}
\end{figure}

The relation of this problem to percolation can be understood
by considering very large $H$. 
In this limit, $\sper\rightarrow 0$, and within the metallic constituent 
the in-plane current prefers to flow in the $\spar$ direction.  
For large enough volume fraction $\psc$ of the $S$ constituent,
and small enough volume fraction $\pins$ of $I$,
there always exist current paths between $S$ grains that are
everywhere parallel to the high-conductivity direction in $M$
[cf.\ Fig.\ \ref{fig:percolation}(a)].   
Therefore, in this regime, 
at large $H$ the effective composite conductivity (and hence resistivity) 
saturates at some finite value.  In the opposite case 
(small $\psc$, large $\pins$), the current
will be forced in some regions to flow in the low conductivity direction 
[cf.\ Fig.\ \ref{fig:percolation}(b)].  In this case, 
the macroscopic or effective conductivity of the system will be
proportional to $\sper$;
hence, the effective resistivity $\rho_e$ satisfies
$\rho_e \sim 1/\sper \sim H^2$, i.e., $\rho_e$ will never saturate.

Next, we qualitatively discuss the expected ``phase diagram'' of this
composite.   The effective conductivity $\hat\sigma_e$ of the 2D mixture is 
a $2\times 2$ tensor
(like the effective 2D resistivity $\hat \rho_e =  \hat \sigma_e^{-1}$),
whose components depend on the applied magnetic field.
If the in-plane microstructure is isotropic, 
then the principal axes of $\hat \sigma_e$ are determined by
${\bf B}$.  At large $H$, both $\sigma_{e,\|}$
and $\sigma_{e,\perp}$ 
can either decrease as $H^{-2}$ for arbitrarily large $H$, 
or else saturate at some finite value.
In principle, there could be a region in the phase diagram where
the resistivity saturates along one principal axis, but not along the other.
But we expect no such region in an infinite system.
This conclusion, as well as the location of the critical line, follow from
the relation between the present problem and the anisotropic percolation
model in 2D, as we now explain.

Anisotropic percolation is usually defined in terms of a random resistor
network (RRN), where the bond occupation probability depends on
the bond orientation. In our 2D network model those probabilities
will be chosen as $p_\parallel=p_M+p_S$ for bonds along {\bf B},
the high conductivity direction of the $M$ constituent,
and $p_\perp=p_S$ for bonds
perpendicular to {\bf B}, the low conductivity direction. For
an infinite network, the percolation threshold for both directions
occurs when \cite{sykes63}
\begin{equation}
\label{anisotropicPT}
1=p_\|+p_\perp=2p_S+p_M\Rightarrow\pins=\psc.
\end{equation}
This threshold separates the regimes of saturating and non-saturating
magnetoresistance. It
agrees with a prediction based on the 
effective medium approximation (EMA), \cite{BergPRLsub,BergmanPRB2001}
as well as with
numerical results below.
In a composite with less symmetry between the $I$ and $S$ constituents,
the threshold will usually differ from $p_I=p_S$.

\begin{figure}[h]
\centerline{ \epsfig{file=\figdir 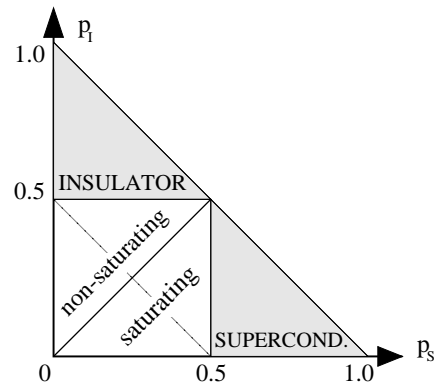,width=2.2in} }
\vskip0.5pc
\caption[]
{
Schematic phase diagram for an $M/I/S$ composite
with random columnar microstructure.  
$\pins$ and $\psc$ are the volume fractions of $I$ and $S$;
the $M$ volume fraction is $p_M=1-\pins-\psc$.   
In the shaded regions, either the $S$ or the $I$ constituent
percolates.  The critical line $\pins=\psc$ separates regions 
with saturating and non-saturating magnetoresistance.  The dotted line
corresponds to percolation of the metallic constituent alone
($p_M>0.5$ or $\pins+\psc<0.5$); no critical behavior appears to 
be associated with this line.
}
\label{fig:diagram}
\end{figure}

The predicted phase diagram is shown in Fig.~\ref{fig:diagram}.
For $\psc>0.5$, the $S$ constituent percolates, hence the composite
is a perfect conductor. Similarly, if $\pins>0.5$ then the
combination of $M$ and $S$ constituents does not percolate, hence
the composite is a perfect insulator. These behaviors are independent
of how much of the $M$ constituent is present.
The remainder of the phase diagram is divided into saturating and 
non-saturating regimes by the line
$\pins=\psc$.   In both regimes, our numerical results indicate that it is 
irrelevant whether the $M$ constituent percolates by itself 
(as in the area below the dotted line), or whether only the 
aggregate of $M$ and $S$ constituents percolates.

\newcommand{\gs}{g_{sat,i}}
\newcommand{\gns}{g_{non-sat,i}}
\section{Critical exponents and scaling properties}

The critical behavior on approaching the critical line
is described by a number of critical exponents.  One of these governs the 
two correlation lengths,
$\xi_\|$ and $\xi_\perp$, which diverge at the critical line.  But we
expect that the divergence of both will be governed by {\em the same}
critical exponent $\nu$, because their ratio $\xi_\|/\xi_\perp$ is determined 
by some function of the constituent volume fractions which 
remains finite on crossing the critical line. \cite{note1,sarychev79}
Thus we write
\begin{equation}
\label{def:xi}
\xi_\| \sim \xi_\perp \sim (\pins-\psc)^{-\nu}.
\end{equation}
Moreover, since the correlation length is just a geometrical property 
of a percolating system, it should have the same value $\nu_{a}$
as in anisotropic percolation, i.e.,
$\nu = \nu_{a}$.  An earlier renormalization group analysis predicts
that anisotropic percolation is governed by the isotropic 
fixed point. \cite{anisotropic}  We therefore expect
that $\nu_{a}$ has the same value
as in 2D isotropic percolation, which has been shown
both analytically\cite{nijs} and numerically\cite{derrida} to equal 4/3.   
However, a similar proof has not been given for anisotropic
percolation; \cite{anisotropic_nu} moreover, some time ago it was reported that
$\nu$ does depend on anisotropy. \cite{shapiro}
Our numerical results below give 
$\nu = 4/3 \pm 0.02$, independent of anisotropy.

The magnetoresistance is described by other critical indices.
At any point such that $\pins<\psc$ 
(the region below the critical line in Fig.~\ref{fig:diagram}), 
the effective conductivity saturates at some finite value
($i=\|$ or $i=\perp$):
\begin{equation}
\label{def:s}
\gs \equiv \lim_{H \rightarrow \infty} \sigma_{sat, i}(H).
\end{equation} 
As we approach the critical line (say, along the line $p_M =$ const), 
$\gs$ must tend to 0,
since on the other side of that line the conductivity vanishes at large $H$ as 
$H^{-2}$. Hence we can introduce the critical exponents $t_{\|}$ and $t_{\perp}$
according to
\begin{equation}
\label{def:gs}
\gs \sim (\psc-\pins)^{t_i}.
\end{equation}
In the region $\pins>\psc$, the effective conductivity is
proportional to $1/H^2$. Therefore, we can define
the finite limiting value
\begin{equation}
\label{def:gns}
\gns \equiv \lim_{H \rightarrow\infty} H^2\sigma_{non-sat,i}(H).
\end{equation}
Since $\gns$ must diverge as the critical line is approached, we can define
the critical exponents $s_{||}$ and $s_{\perp}$, according to
\begin{equation}
\label{def:t}
\gns \sim (\pins-\psc)^{-s_{i}}.
\end{equation}
Just as in the case of $\nu$, we expect $s_{\|}= s_{\perp}$, 
$t_{\|}=t_{\perp}$.
We might expect that the actual values of $t$ and $s$ are
determined, once again, by the connection to anisotropic percolation.
However, this connection is more subtle than for $\nu$, 
because $t$ and $s$ refer to different
physical quantities in the present problem, than in anisotropic
percolation.  
Nonetheless, it is reasonable to expect that both problems
belong to the same universality class, implying
$s_i = t_i$ = 1.30, the values of the two conductivity exponents in
conventional 2D percolation problems. \cite{t2D,YDelta,t2Dnew}
Note also that EMA yields $s_i=t_i=1$, \cite{BergPRLsub,BergmanPRB2001}
and that continuum composites sometimes belong to a different
universality class of percolation, with different values of
the critical exponents, depending upon the microstructural details.
\cite{FengHalperinSenPRB87}


Finally, consider systems exactly at the percolation threshold, i.e., at
$p_I = p_S$.  Any finite-size sample inevitably falls into either the
percolating class, with $\rho_e \sim H^0$ as $H \rightarrow \infty$, or the
non-percolating class, with $\rho_e \sim H^2$.  
On increasing the size of the system,
one finds that the field where a saturating sample achieves saturation also
increases.  The same size effect also holds for non-saturating
samples: the characteristic field at which $\rho_e$ starts to vary as $H^2$
increases with the size of the system.  Below that field, or at any field
in the limit of a system of infinite size, the behavior of the
magnetoresistance is described by yet another critical exponent
$\delta_i$, defined by
\begin{equation}             
\label{def:delta}  
\rho_{e,i}\sim|H|^{\delta_i}.
\end{equation}     
The EMA analysis\cite{BergPRLsub,BergmanPRB2001} yields $\delta_i=1$.

In fact, the result $\delta_i = 1$
also follows from a duality argument, if one assumes that
$\delta_\| = \delta_\perp$. The duality principle\cite{SSP} gives
$$1=\sigma_{e,\|}\left(
	\sigma_{M\|},\sigma_{M\perp},\sigma_S,\sigma_I
	\right)
\sigma_{e,\perp}\left(
	\frac{1}{\sigma_{M\perp}},\frac{1}{\sigma_{M\|}},
	\frac{1}{\sigma_S},\frac{1}{\sigma_I}
	\right).$$
It immediately follows that the anisotropic percolation thresholds
for the two directions $\parallel$, $\perp$ must coincide, for
otherwise the singular behaviors of $\sigma_{e,\|}$ and $\sigma_{e,\perp}$
would not cancel, as required by this equation. It also follows,
rigorously, that $t_\parallel=s_\perp$ and $t_\perp=s_\parallel$.
Using the homogeneity of $\sigma_{e,\|}$, $\sigma_{e,\perp}$
as functions of the various constituent conductivities, and noting
that $\sigma_I=0$ and $\sigma_S=\infty$, we can rewrite this equation as
$$\frac{\sigma_{M\perp}}{\sigma_{M\|}}=
\sigma_{e,\|}\left(1,\frac{\sigma_{M\perp}}{\sigma_{M\|}},\infty,0\right)
\sigma_{e,\perp}\left(1,\frac{\sigma_{M\perp}}{\sigma_{M\|}},0,\infty\right).$$
Note that the relation $p_I=p_S$ for the percolation threshold follows
rigorously from this equation if the microstructure is invariant
under the interchange of the $S$ and $I$ constituents.
At the percolation threshold, and when $|H|\gg 1$, this reduces to
$$\frac{1}{H^2}\propto\frac{1}{|H|^{\delta_\parallel+\delta_\perp}}
\Rightarrow\delta_\parallel+\delta_\perp=2.$$
The assumption that $\delta_\| = \delta_\perp$, and hence the final
result $\delta_\| = \delta_\perp=1$, are supported by the physical
picture of the anisotropic percolation process, and is consistent with
the expectation that $t_\parallel=t_\perp$ and $s_\parallel=s_\perp$.
These equalities also follow if a simple scaling description applies
to both $\sigma_{e,\|}$ and $\sigma_{e,\perp}$ with {\em the same
scaling variable}. Thus, for an $M/I/S$ columnar composite precisely
at the critical composition (i.e., on the phase boundary line between
saturating and non-saturating magnetoresistance), we expect to
find
$$\sigma_{e,\|} \sim \sigma_{e,\perp} \sim\frac{1}{|H|},\;\;\;
\rho_{e,\|} \sim \rho_{e,\perp} \sim |H|\Rightarrow\delta=1.$$

Near the transition, where $|p_I-p_S|\ll 1$ and
$\sigma_{M,\perp}/\sigma_{M,\|}\cong 1/H^2\ll 1$,
we expect that a scaling description of the critical behavior is applicable.
In view of the preceding discussion, that description can be formulated
as follows:
\begin{eqnarray}
\frac{\rho_{e,i}}{\rho_{M,\|}}&\cong&{\rm sign}\Delta p\,|\Delta p|^{-t}
 F_i(Z);
\label{def:F}\\
\Delta p&\equiv& p_I-p_S,\;\;\;
 i=\|,\,\perp,\;\;\;
 Z\equiv |H||\Delta p|^t\,{\rm sign}\Delta p.
\nonumber
\end{eqnarray}
The scaling functions $F_i(Z)$ should have the following asymptotic
forms for extreme values of the scaling variable $Z$:
\begin{equation}
F_i(Z)\sim\left\{\begin{array}{lll}
-Z^0 & {\rm for}\;Z\ll-1,& {\rm i.e.}\;p_S>p_I,\\
Z^2 & {\rm for}\;Z\gg 1,& {\rm i.e.}\;p_S<p_I,\\
Z & {\rm for}\;|Z|\ll 1,& {\rm i.e.}\;p_S\stackrel{>}{\scriptstyle <}p_I.
\end{array}\right.
\label{F_i_Z}
\end{equation}
The first two lines in this expression follow from the fact that
we expect to have $\rho_{e,i}\sim H^0$ in the saturating regime
and $\rho_{e,i}\sim H^2$ in the non-saturating regime. The
third line results from the necessity to cancel the dependence
of $\rho_{e,i}$ upon $\Delta p$ when $\Delta p\rightarrow 0$.
This kind of behavior was already found previously using
EMA, \cite{BergPRLsub,BergmanPRB2001} where the incorrect value
$t=1$ was found, as is usual when that approximation is invoked.
Note that, even though $F_i(Z)$ has a different
analytic form for large $Z$, depending on the sign of $Z$ or $\Delta p$,
these functions are expected to vary smoothly when $Z$ passes
through 0. 

Finally, note that duality is a general symmetry property of 2D systems,
and does not require that the $I$ and $S$ inclusions have similar
shapes or spatial distributions. Thus, the conclusions regarding
values of $s$, $t$, $\delta$ are valid even if the line of
critical compositions differs from the simple straight line
$p_S=p_I$, which is valid only for the case where the $I$ and $S$
constituents appear in the composite in a symmetric fashion.

\section{Results}

We carried out calculations on simple-square-lattice RRN's, choosing the
resistors in accordance with the model described above.  
The calculation was done using the Y-$\Delta$ bond elimination 
algorithm. \cite{YDelta}  Besides being very efficient in 2D, 
this algorithm allows the inclusion of both 
perfectly insulating and perfectly conducting 
bonds without any approximations.
This technique is much more efficient than the techniques available
for 3D networks. That is why the results obtained here for columnar
systems are much more detailed and accurate than the results obtained
previously for 3D isotropic systems by simulations of 3D random
network models. \cite{SarBergStrelPRB93}

The networks for our results were constructed as follows:
each bond was independently and randomly 
chosen to be insulating, perfectly conducting, 
or metallic, with appropriate probabilities.    A metallic bond was
assigned an appropriate conductance, depending on whether it was oriented
parallel or perpendicular to the magnetic field.   To calculate
$\gs$, as can be seen from Eqs.\ (\ref{2Dsigmas}) and (\ref{def:s}),  
the conductances of the constituents should be taken as
\begin{equation}
\label{Ssigmas}
\sins=0 \text{ , } 
\ssc=\infty \text{ , } \
\spar=1 \text{ , } 
\sper=0.
\end{equation}
But in order to calculate 
$\gns$, one needs to multiply all the conductances by $H^2$ and take the limit
$H\rightarrow \infty$, which leads to
\begin{equation}
\label{NSsigmas}
\sins=0 \text{ , } 
\ssc=\infty \text{ , } \
\spar=\infty \text{ , } 
\sper=1.
\end{equation}

The definitions (\ref{def:gs}) and (\ref{def:t}) for the exponents $s$ and $t$
are exactly valid only for an infinite system.
Instead of using these definitions directly, we adopted the
finite-size scaling approach, as described below. 
For a finite-size system generated {\em exactly at} 
the percolation threshold $\pins=\psc$,
the system size $L\times L$ is always smaller than the (infinite) correlation 
length $\xi$. From Eq.\ (\ref{def:xi}) it follows that such a system
behaves like a system at another volume fraction satisfying
$|\pins-\psc| \sim L^{-1/\nu}$.  
Therefore, the average of $\gs$ or $\gns$ over many such systems
should scale as $\left< \gs\right> \sim L^{-t_i/\nu}$ or 
$\left<\gns\right>\sim L^{s_i/\nu}$. (Due to the appearance of
percolating samples, the latter average is actually 
infinite. A finite result is obtained by averaging
only over non-percolating samples.  This procedure 
does not change the scaling form, because {\em at the percolation threshold}
the fraction of samples which are non-percolating is 
asymptotically size-independent, as discussed below).

\begin{figure}
\epsfig{file=\figdir 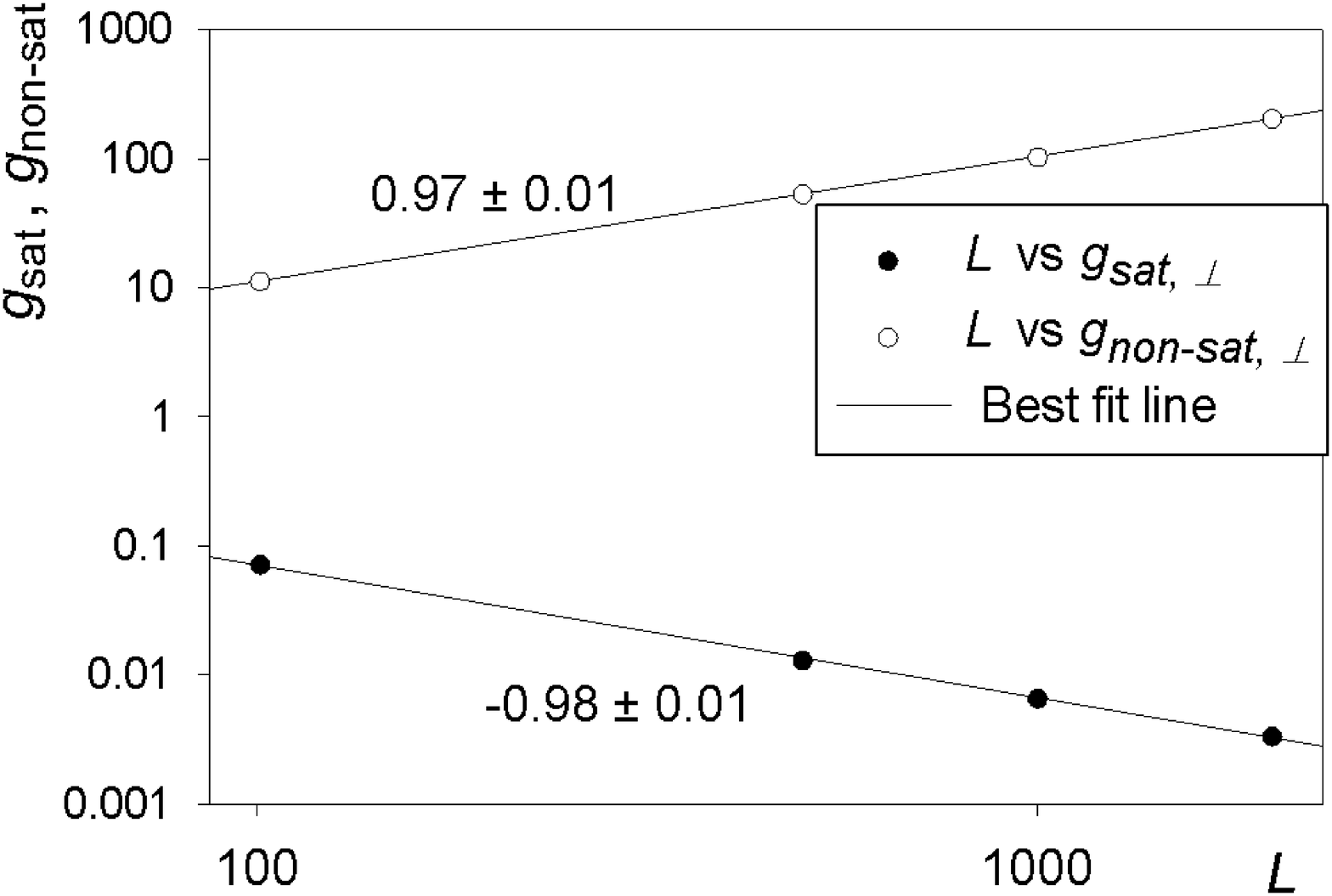, width=3.in}
\vskip0.3pc
\caption[]
{
Calculated $\left< \gs \right>$ and $\left< \gns \right>$
[cf.\ definitions (\ref{def:s}) and (\ref{def:gns})] plotted on a log-log
plot versus linear size $L$ of the RRN for $p_M=0.2$ and
$\pins=\psc=0.4$.
Error bars are smaller than the dot sizes.
The slopes of the two lines on this plot equal $t_i/\nu$ and $-s_i/\nu$,
and are consistent with the value 0.975, corresponding to $t=s=1.30$
and $\nu=4/3$.
}
\label{fig:scaling}
\end{figure}

Using RRN's with $L$ ranging from 100 to 2000 and $p_I=p_S$, 
we estimated $s_i$ and $t_i$ for $p_M=0.2$, $0.5$, $0.8$, and, with 
less accuracy, for several other values of $p_M$.
(At $p_M=0.8$ we could estimate only $t_\|$ and $s_\perp$, since these systems 
usually percolated only in one direction but not in the other.)
The finite size scaling assumption is supported very well, as demonstrated 
in Fig.~\ref{fig:scaling} for $p_M=0.2$. 
Assuming $\nu=4/3$, we find for all values of $p_M$ studied that 
$t_{||} =t_{\perp} =s_{||}=s_{\perp}=1.30\pm 0.02$.
This value supports the hypothesis that the problem belongs to the same 
universality class as both isotropic and anisotropic percolation in 2D.

\begin{figure}
\centerline{\epsfig{file=\figdir 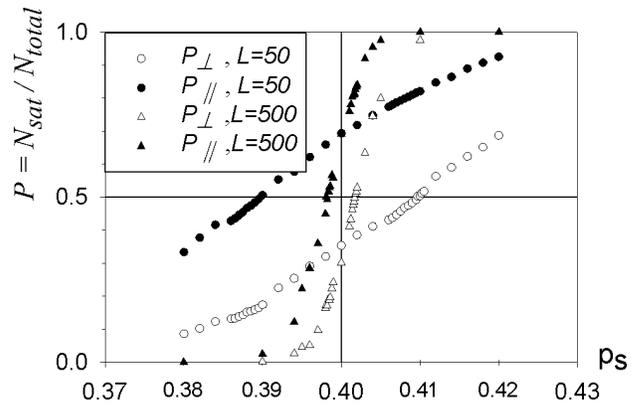,width=3.2in}}
\vskip2pc
\caption[]{
Fraction $P_i\equiv N_{sat}/N_{total}$ of samples which
saturate in the direction $i$, plotted versus
$S$ volume fraction $\psc$,
at fixed $M$ volume fraction $p_M=0.2$, for two different sizes $L$.
The vertical line denotes the percolation threshold in
an infinite system, as predicted by Eq.\ (\protect\ref{anisotropicPT}).
For a finite system, the ``percolation threshold'' is
the point where one half of the samples percolate: $P_i=0.5$.
Such thresholds are unequal in the directions parallel and
perpendicular to ${\bf B}$, but
approach the same infinite-size value,
${\psc}_0=0.4$,
as $L\rightarrow \infty$.
}
\label{fig:probabilistic}
\end{figure}

The assumption that $\nu=4/3$ can be tested for our problem 
by considering percolation in finite-size systems.  For such systems,
the ``percolation threshold'' in the direction $i$
is naturally defined as the value of $p_S$ for which exactly
one half of the systems percolate in that direction:
$P_i\equiv N_{sat,i}/N_{total}=1/2$, where $P_i$ is the probability
of finding saturating behavior in the direction $i$.
Fig.~\ref{fig:probabilistic} then shows
that the percolation threshold depends on the system size $L$. In
fact, a finite system has
two ``percolation thresholds'', ${\psc}_{0,\perp}$ and ${\psc}_{0,\|}$, located
respectively above and below the percolation threshold of an infinite system,
and approaching $p_I$ as $|{\psc}_{0,i} - p_I|\sim L^{-1/\nu}$
with increasing $L$. \cite{PercThresh}
We have checked these hypotheses numerically by calculating 
$| {\psc}_{0,\perp} - {\psc}_{0,\|} |$ for four linear 
sizes $L$ at $p_M=0.2$ and $0.5$, 
and verified that $\nu=4/3 \pm 0.02$, independent of anisotropy. 

Fig.\ \ref{fig:probabilistic} also makes clear that the
fraction of samples which are non-percolating
is size-independent at the percolation threshold.  
Specifically, Fig.~\ref{fig:probabilistic} shows
the fraction of percolating samples of a given size vs.\ concentration $p_S$.
The lines corresponding to different sample sizes $L$ intersect
at the percolation threshold $p_I=p_S$ (vertical line).
Thus, as $L \rightarrow \infty$,
the fraction of the samples percolating at $p_I=p_S$ approaches
a certain limiting value determined by the other parameters of the system, 
namely, the direction of percolation and the concentration $p_M$ of the metal.
This limiting value can be related to the spanning probability in the case
of {\em isotropic} percolation in a system of rectangular shape, 
which was found to depend on the aspect ratio
of the rectangle. \cite{SpanningProbability}
In the present paper, we only consider systems 
with the aspect ratio equal to unity; however, the {\em correlation lengths}
are different in the two directions. Thus, $\xi_\perp / \xi_\|$
may be thought of as an effective aspect ratio for the present problem.
 
\begin{figure}[t]
\centerline{\epsfig{file=\figdir 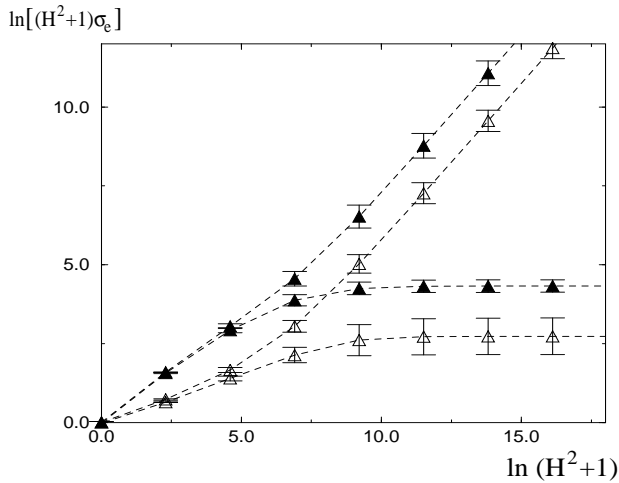,width=3.5in}}
\vskip1pc
\caption[]{
$\ln [(H^2+1) \sigma_{e,i}]$ vs. $\ln (H^2+1)$ for systems with
$L=100$ and $p_M=0.5$ at the percolation threshold $p_I=p_S$. 
Averaging over different realizations is performed separately for
saturating and non-saturating samples.
Open symbols correspond to $\sigma_{sat,\perp}$ and $\sigma_{non-sat,\perp}$
and the filled symbols to $\sigma_{sat,\|}$ and $\sigma_{non-sat,\|}$.
For values of $H$ within and below the ``critical'' range
of fields (as defined in the text), the conductivity in a given direction
does not depend on whether or not it saturates at stronger fields $H$.
}
\label{fig:ThirdIndexSameLine}
\end{figure}

\begin{figure}[t]
\centerline{\epsfig{file=\figdir 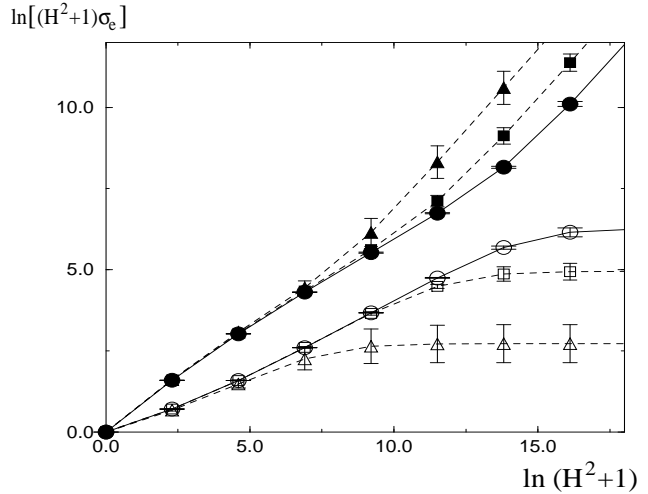,width=3.6in}}
\vskip1pc
\caption[]{
$\ln [(H^2+1) \sigma_{e,i}]$ vs. $\ln (H^2+1)$ for systems of size
$L=$ 100 (triangles), 1000 (squares) and 4000 (circles), with 
$p_M=0.5$, $p_I=p_S=0.25$. Open symbols correspond to $\sigma_{e,\perp}$
and the filled symbols to $\sigma_{e,\|}$. The lines are drawn as
guides for the eye.
At large fields, all systems exhibit
either saturating ($H^2 \sigma_i\sim H^2$) or non-saturating 
($H^2 \sigma_i\sim 1$) magnetoresistance. 
In the ``critical'' range of fields,
clearly seen for the larger systems, the magnetoresistive behavior is
given by Eq.\ (\ref{def:delta}) with 
$\delta_\perp = \delta_\| = 1.00 \pm 0.07$.
Note that, for the sake of visual clarity, we include only those systems
in which $\sigma_{e,\perp}$ does not saturate and
$\sigma_{e,\|}$ does saturate [cf. Fig.~\ref{fig:ThirdIndexSameLine}].
}
\label{fig:ThirdIndex}
\end{figure}

In order to test the scaling behavior at finite values of $H$ and
$\Delta p$, we also simulated random networks at $p_S=p_I$
with large but finite values of both $H$ and $L$. 
Since the RRN's required for such calculations
contain bonds with widely different
conductivities [cf.\ Eq.\ (\ref{2Dsigmas})], special care must be taken
when performing these calculations on large systems. \cite{note:CPP}
For that reason we wrote special code for storing numbers as large (small) 
as roughly $10^{\pm 6\times 10^8}$.
Using this code, we studied RRN's of size up to
$4000\times 4000$ in fields ranging from 0 up to $H^2=10^8$.
One useful way to
exhibit those results is as plots of $\ln[(H^2+1)\sigma_{e,i}]$
vs.\ $\ln(H^2+1)$ for different values of the linear size $L$ --- see
Figs.\ \ref{fig:ThirdIndexSameLine} and \ref{fig:ThirdIndex}.
Fig.~\ref{fig:ThirdIndex} shows $\sigma_{e,\|}$ and $\sigma_{e,\perp}$
vs. magnetic field in systems with sizes ranging from 100 to 4000.
For the larger systems, one can clearly see a ``critical'' range of
fields, in which the magnetoresistance is consistent with the scaling
form of Eq.\ (\ref{def:delta})
with $\delta_\perp = \delta_\| = 1.00 \pm 0.07$.
For fields within and below that critical range,
the behavior of $\sigma_{e,i}$
is independent of $L$, and is also independent of
whether its value does or does not saturate at higher fields,
as demonstrated in Fig.~\ref{fig:ThirdIndexSameLine}.
However, the intercepts of those linear dependencies
are different for the different directions:
$\sigma_{e,i} \sim (|H|- H_{0,i})^{\delta_i}$, with positive $H_{0,\perp}$
and negative $H_{0,\|}$. This is due to the fact that the system always
conducts better in the direction parallel to the applied magnetic field,
as can be surmised from Eq.\ (\ref{2Dsigmas}).
Figs.\ \ref{fig:ThirdIndexSameLine} and \ref{fig:ThirdIndex}
bear out the forms hypothesized earlier for $F_i(Z)$---see
Eq.\ (\ref{F_i_Z}). In particular, Fig.\ \ref{fig:ThirdIndexSameLine}
bears out the expectation that, for small $|Z|$, $F_i(-Z)=-F_i(Z)$,
and thus that the scaling functions are smooth at $Z=0$.
Figs.\ \ref{fig:ThirdIndexSameLine} and \ref{fig:ThirdIndex} also
show that both $\sigma_{e,\|}$ and $\sigma_{e,\perp}$  are
independent of $L$ when $|H|\ll L^{t/\nu}$, in agreement with
Eqs.\ (\ref{F_i_Z}) and (\ref{rho_e_L}) below.

In order to find the form of the scaling functions $F_i(Z)$ 
[defined by Eq.\ (\ref{def:F})] from these numerical results,
it is convenient to also
invoke the finite-size-scaling hypothesis:
\cite{AharonyStauffer92} In a system of finite
linear size $L$, when the correlation length $\xi$ satisfies
$\xi\gg L$, $|\Delta p|$ should be replaced by $C_1/L^{1/\nu}$
in all the expressions of Eq.\ (\ref{def:F}):
\begin{equation}
\frac{\rho_{e,i}}{\rho_{M,\|}}\cong L^{t/\nu}{\rm sign}\Delta p\,F_i(Z),\;\;\;
Z\equiv{\rm sign}\Delta p\,|H|/L^{t/\nu}.\label{rho_e_L}
\end{equation}
Note that the constant $C_1$ has been absorbed into the definitions of
$Z$ and $F_i(Z)$. The sign of $\Delta p$, 
appearing in Eq.\ (\ref{rho_e_L}), should now be determined from the actual
behavior (i.e., percolating vs.\ non-percolating, or saturating
vs.\ non-saturating, in the direction $i$)
of {\em each particular sample}.

\begin{figure}[t]
\centerline{ 
 \epsfig{file=\figdir 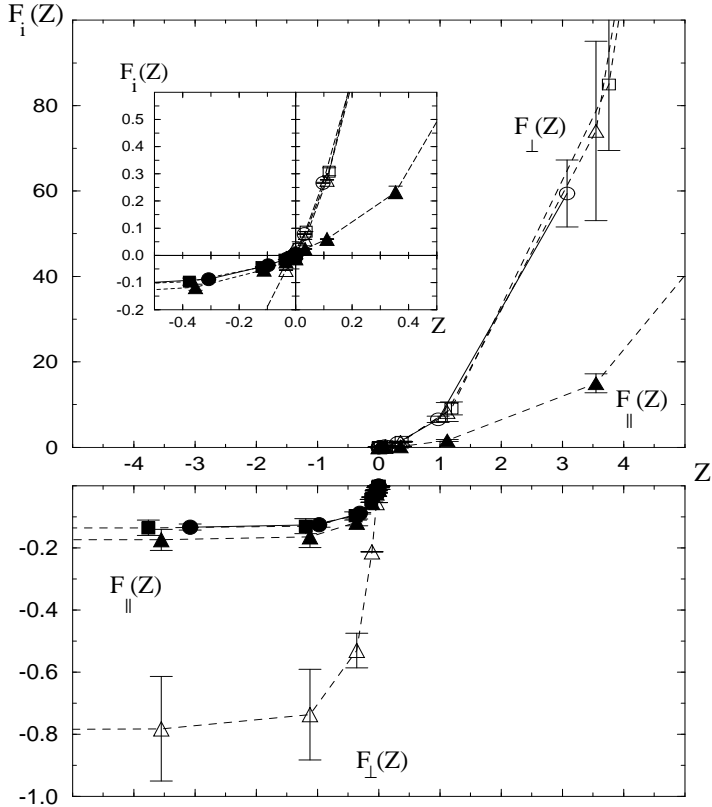,width=4.25in}
}
\vskip-1pc
\caption[]{Plots of $F_i(Z)\equiv
{\rm sign}\Delta p\,L^{-t/\nu}\rho_{e,i}/\rho_{M,\|}$
vs.\ $Z\equiv{\rm sign}\Delta p\,|H|L^{-t/\nu}$ for networks with $p_M=0.5$ and
various large but finite values of $H$ and $L$. The top frame shows
$F_i(Z)$ in the non-saturating regime, when $Z>0$, while the
bottom frame shows $F_i(Z)$ in the saturating regime, when $Z<0$.
The shapes of the points correspond to the value of $L$, as in
Fig.\ \ref{fig:ThirdIndex}. Open symbols correspond to $F_\perp(Z)$,
while filled symbols correspond to $F_\|(Z)$. The horizontal lines surrounding
the symbols are parts of error bars. Inset: expanded
view of $F_i(Z)$ for small values of $|Z|$ and both signs of $Z$,
showing the smooth linear behavior of $F_i(Z)$.
}
\label{scaling}
\end{figure}

The plots of $L^{-t/\nu}\rho_{e,i}/\rho_{M,\|}$ vs.\ $|H|L^{-t/\nu}$,
shown in Fig.\ \ref{scaling}, clearly demonstrate that
the results obtained for $\rho_{e,i}$, using different values
of $H$ and $L$, collapse onto a single curve (for a given $i$) 
when scaled in
accordance with Eq.\ (\ref{rho_e_L}). These figures constitute
a quantitative graphical representation of the scaling functions
$F_\|(Z)$ and $F_\perp(Z)$ for the special case $p_M=0.5$.
Note that these two scaling functions appear to have a similar shape,
up to a constant coefficient. That is qualitatively consistent with the
results of EMA, which lead to \cite{BergmanPRB2001}
$$\frac{F_\perp(Z)}{F_\|(Z)}=
\left(1+p_M\over 1-p_M\right)^2.$$
In the case of the systems featured in Fig.\ \ref{scaling},
where $p_M=0.5$, this ratio should be 9 according to EMA. By contrast, the
simulation results plotted in that figure indicate that this
ratio is between 4.5 and 5. As stated earlier, such quantitative
discrepancies should come as no surprise, in view of the known
deficiencies of EMA in the critical region near a percolation threshold.

\section{Discussion}
We have demonstrated that the phase diagram of an
$M/I/S$ random columnar composite in an in-plane magnetic field exhibits
a critical line between saturating and non-saturating regimes
of magnetoresistance.
The behavior near this critical line was interpreted
using an anisotropic percolation model.
The critical exponents $t=s$ and $\nu$ were found, numerically,
to be the same as in the more conventional 2D isotropic
random percolating networks. Also found were the exponents
which govern the dependence on magnetic field in the critical
region. The existence of scaling behavior in that region was
confirmed and the forms of the scaling functions were obtained numerically.

The critical line discussed here could
be studied experimentally using a doped semiconductor film as 
the $M$ constituent,
with a random collection of etched perpendicular holes as the
$I$ constituent, and a random collection of perpendicular
columnar inclusions, made of a high conductivity normal metal,
playing the role of $S$.
Extremely low temperatures or very clean single crystals
would not be required in order to observe this critical line.
What would be necessary is a large contrast at each stage of
the following chain of inequalities
$\rho_S\ll\rho_M\ll H^2\rho_M\ll\rho_I$. 
If Si-doped GaAs is used as the $M$ host, with a negative charge
carrier density of $1.6\times 10^{18}$ cm$^{-3}$ and a mobility
$\mu=2500$ cm$^2/$V\,s at a temperature of 90\,K,
as in the experiment described in Ref.\
\onlinecite{TornowWeissKlitzingEberlBergStrelPRL96},
then a magnetic field of 40\,Tesla
would result in $H=-10$. Such a material would have an Ohmic
resistivity of $1.6\times 10^{-3}\,\Omega$\,cm, about 1000
times greater than that of Cu. Thus, using Cu for the $S$
inclusions and etched holes for the $I$ inclusions, all the above inequalities
could be satisfied without difficulty.

This is apparently the first experimentally accessible and technologically
promising system in which anisotropic percolation is relevant.
Our numerical results for the exponents $\nu$, $t$, $s$, and $\delta$
are consistent with the assumption that this problem 
belongs to the same universality class as the usual 2D isotropic
percolation problem, and confirm that $\nu$ is independent of anisotropy.

\acknowledgements

We are grateful to Y. Kantor for providing us with the idea 
regarding the $L$ dependence of $P_i$ and the appropriate references.
This work was supported in part by NSF Grant DMR97-31511, and by
grants from the
US-Israel Binational Science Foundation and
the Israel Science Foundation.


\begin{thebibliography}{5}
\bibitem{note:CPP}
Indeed, application of the Y-$\Delta$ algorithm\cite{YDelta} to such a system
leads to generation of intermediary bonds with rapidly growing
exponents, both positive and negative. It is important to keep track of such 
very large and very small numbers, especially in a system that is
close to a percolation threshold:
a single replacement of a very weak bond by a bond of zero conductance
changes the topology of the network, possibly
driving it from percolating to non-percolating.
\end{thebibliography}
\end{document}